\documentclass[
aps,%
12pt,%
final,%
notitlepage,%
oneside,%
onecolumn,%
nobibnotes,%
nofootinbib,% 
superscriptaddress,%
noshowpacs,%
centertags]%
{revtex4}
\RequirePackage[utf8]{inputenc}   %for home
\RequirePackage{amssymb,amsmath}
\RequirePackage{eufrak}
\RequirePackage{oldlfont}
\RequirePackage{mathtext}
\RequirePackage{bm}
\RequirePackage{array}
\RequirePackage{longtable}
\RequirePackage{graphicx}
\RequirePackage{calc}

\begin{document}
%\selectlanguage{russian}

\title{Spectral Survey \\ of the Star Formation Region DR21OH \\ in the 4 mm Wavelength Range}%  §¡š¥­š¥ ­  áâà®ªš ®áãé¥áâ¢«ï¥âáï ª®¬ ­€®© \\

\author{\firstname{S.~V.}~\surname{Kalenskii}}
% €¥áì à §¡š¥­š¥ ­  áâà®ªš ®áãé¥áâ¢«ï¥âáï  ¢â®¬ âšç¥áªš š«š ª®¬ ­€®© \\
\email{kalensky@asc.rssi.ru}
\affiliation{%
Lebedev Physical Institute, Astro Space Center,\\ 84/32 Profsoyuznaya st., Moscow, GSP-7, 117997, Russia
}%
\author{\firstname{E.~A.}~\surname{Mikheeva}}
\email{mikheeva-ekaterina99@yandex.ru }
\affiliation{%
Lebedev Physical Institute, Astro Space Center,\\ 84/32 Profsoyuznaya st., Moscow, GSP-7, 117997, Russia
}
\affiliation{%
Sternberg Astronomical Institute, Moscow State University, \\ Universitetsky pr., 13, Moscow 119234, Russia
}%

%\date{\today}
%\today ¯¥ç â ¥â c¥£®€­ïè­¥¥ çšá«®

\begin{abstract}
The results of a spectral survey of the region of massive star formation DR21OH in the 4-mm wavelength range are presented. Sixty-nine molecules and their isotopologues have been detected, ranging from simple diatomic or triatomic species such as SO, SiO and CCH, to complex organic molecules such as CH$_3$OCHO or CH$_3$OCH$_3$. The obtained results qualitatively repeat the results of the survey of the same source at 3~mm. The inventories of molecules found at 3mm and 4mm overlap to a great extent. However, at 4 mm we found a number of species that have no allowed transitions in the 3-mm wavelength range, e.g. DCN, DNC, or SO$^+$. The bulk of the molecules detected at 4~mm are those that are common for dense cores, e.g., HC$_3$N or CH$_3$CCH, but some of the detected species are typical for hot cores. The latter include complex organic molecules CH$_3$OCHO, CH$_3$CH$_2$OH, CH$_3$OCH$_3$, etc. However, the detected emission of these molecules probably arises in a gas heated to 30 K only. Nine molecules, including complex species CH$_3$C$_3$N, CH$_3$CH$_2$CN, CH$_3$COCH$_3$, were found by spectral line stacking. This demonstrates the prospects of the method in the study of molecular clouds.
\end{abstract}

\maketitle

\section{INTRODUCTION}

The high-mass star-forming region DR21OH is located within a giant molecular cloud in the Cygnus X complex. The distance to the complex, obtained from the trigonometric parallaxes of H$_2$O and CH$_3$OH masers, is 1.4 kpc~\cite{rygl12}. DR21OH is a part of a massive dense cloud elongated in the north-south direction for approximately 4 pc  \cite{harvey86,hennemann12}. Within the cloud, there is a chain of Young Stellar Objects (YSOs) also elongated in the north-south direction. In addition to DR21OH, this YSO chain includes a powerful radio continuum source, the bright HII region DR21, which gives the entire cloud its name \cite{harvey86},  as well as the infrared sources W75S-FIR1, W75S-FIR2, and W75S-FIR3. The bolometric luminosity of the DR21OH region is approximately $5\times 10^4$ solar luminosities \cite{harvey77}, and the total mass is about $10^4$ solar masses.

The region is located three angular minutes north of the DR21 HII-region and consists of four main clouds designated DR21OH Main (DR21OH M), DR21OH North (DR21OH N), DR21OH West (DR21OH W), and DR21OH South (DR21OH S), surrounded by molecular gas with a clumpy structure~\cite{mangum91,mangum92}. The DR21OH~M cloud, in turn, consists of two dense cores, MM1 and MM2, whose emission maxima are located 8 seconds apart in the northeast-southwest direction. These cores are observed  both in molecular lines and in the continuum in the mm, submm and IR bands~\cite{padin89,mangum91,mangum92}. Zapata et al. \cite{zapata12} observed DR21OH~M at 1.4 mm wavelength using the SMA interferometer and separated MM1 and MM2 into nine more compact clumps, designated SMA1 -- SMA9. The main physical parameters of MM1 and MM2 were determined by Mangum et al.~\cite{mangum91}. The bolometric luminosity of MM1 is $1.7\times 10^4$ solar luminosities, its mass is 350 solar masses, and its dust temperature is 58~K. The upper limit on the bolometric luminosity of MM2 is $1.3\times 10^4$ solar luminosities; MM2 has a mass of 570 solar masses; the dust temperature is 30 K. The systemic velocity of DR21OH~M is about $-3$~km~s$^{-1}$, but spectral details with LSR velocities about $9$ km s$^{-1}$ are observed toward this cloud. These details are related to a cloud that probably collides with DR21~\cite{dickel78,dobashi19}. According to Mangum et al~\cite{mangum92}, the radial velocities of the MM1 core and several neighbouring clumps are close to the systemic velocity and fall in the interval from -4.4~km~s$^{-1}$ to -2.2~km~s$^{-1}$, while the velocities of the MM2 core and other clumps fall in the interval from 0~km~s$^{-1}$ to -1~km~s$^{-1}$. However, according to the spectra reported in~\cite{minh12}, in the direction of the hot core MM1b (see next paragraph), the lines of $^{34}$SO, HC$_3$N and some other molecules have maxima at velocities around 0 km~s$^{-1}$. %(Dickel et al. 1978; Dobashi et al. 2019). 

An indication of active star formation in DR21OH~M is the existence of strong OH, H$_2$O and CH$_3$OH masers~\cite{genzel77, norris82, batrla88}. In addition, bipolar outflows in CO, CS, SiO, H$_2$CO, H$_2$CS lines have been observed in this region ~\cite{lai03, zapata12, orozco19}. However, despite many indications of an actively proceeding process of high-mass star formation, no compact HII regions are found in DR21OH-M, and the hot cores found by Minh et al.~\cite{minh12} in the MM1 direction and labelled MM1a and MM1b are not as pronounced as the hot cores in sources like Sgr~B2, W51e1/e2, G34.26+0.15, etc. These facts indicate that the protostars in DR21OH~M are at the early stages of evolution. Nevertheless, in this source, we can search for COMs (complex organic molecules), which appear in the gas phase due to various non-thermal desorption mechanisms, as well as due to evaporation of dust mantles in hot postshock gas the lobes of high-velocity outflows. 

A spectral survey of DR21OH at 3 mm wavelengths was carried out by Kalenskii and Johansson (2010)~\cite{kalenskii10a} with the 20-m radio telescope of the Onsala Space Observatory (Sweden). The whole DR21OH~M region fell within the telescope beam. Lines of 78 molecules were detected, most of them typical of dense clouds. The rotational temperatures of most of the detected molecules proved to be about 20 -- 30 K, but the rotational temperature of SO$_2$ was 186 K. An even higher rotational temperature, 252 K, was obtained from methanol lines with upper level energies above 100~K. Some of the detected molecules were COMs - oxygen-containing CH$_3$OH, CH$_3$OCHO, CH$_3$OCH$_3$, nitrogen-containing CH$_3$CN, CH$_3$CH$_2$CN and others.  

The detection of these molecules shows that DR21OH is a promising source to search for COMs. However, a number of molecules important for astrochemistry do not have allowed transitions in the 3-mm wavelength range and must be searched for in other spectral bands. Some molecules --- CH$_2$NH, DC$_3$N, HCCN --- have been found at 3~mm only presumably and their detections should be confirmed or denied. Therefore, when the 4~mm wavelength range receiver appeared at the Onsala Observatory, we carried out a spectral survey of DR21OH in this range as well.

\section{Observations}
The observations were carried out in April and May 2018 on the 20-m millimeter-wave radio telescope of the Onsala Space Observatory (Sweden) in parallel with the spectral survey of the star-forming region W51e1/e2  (project O2017b-04) and are described in detail in~\cite{kalen22}. The spectral range of the survey was 68 -- 88 GHz. The dual beam switching with a beam throw of 11 arcmin in azimuth was used. The main beam width at half-power (HPBW) at 86~GHz was $43''$. The antenna was pointed toward MM1 (R.A.=$20^h39^m01.1^s$, DEC=$42^{o}22'48''$; J2000). The entire DR21OH~M cloud falls within the beam. Pointing errors were checked using SiO maser observations at 86~GHz and did not exceed $5''$. The data were calibrated by applying the chopper-wheel method. The system noise temperature mainly varied between $\sim 200 -- 500$~K, but in the final observing sessions, between $\sim 300$~K and $\sim 1200$~K.
A fast Fourier transform spectrometer consisting of four sections, two sections for each polarisation plane, was connected to the receiver backend. The full analysis bandwidth was 4~GHz. The frequency resolution was 76.294~kHz, corresponding to a velocity resolution of 0.28~km~s$^{-1}$ at 83~GHz. During data processing, the resolution was smoothed to $0.305$~~MHz (1.1~km~s$^{-1}$ at 83~GHz). For 75\% of the survey range, the $1\sigma$ noise level did not exceed 0.003~K, increasing slightly (to 0.006K) toward the edges of the range.

This project served as a test for the new Bifrost control system, which controls the observations in automatic mode.

The data were processed using the Grenoble Observatory's CLASS software package\footnote{https://www.iram.fr/IRAMFR/GILDAS} using the WEEDS extension.

\section{Results}
\begin{figure}
    \centering
    \includegraphics[width=\textwidth]{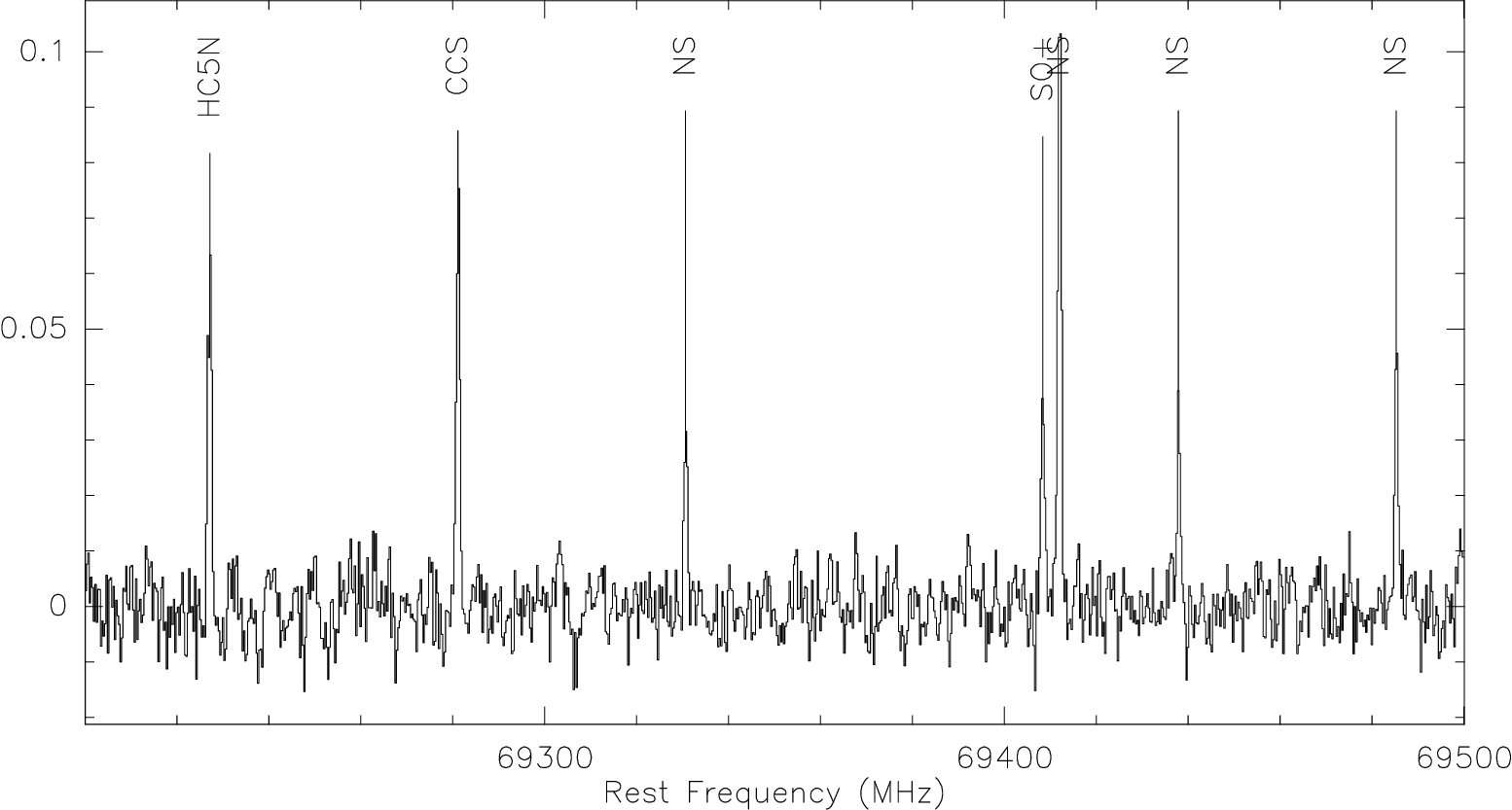}
    \caption{DR21OH spectrum in the frequency range 69200--69500 MHz. The spectrum is fully available in the electronic form.}
    \label{fig:spec}
\end{figure}

Source spectrum in the frequency range 68--88~GHz was obtained. A part of it is reproduced in Fig.~\ref{fig:spec}. Many radio lines of different molecules were detected. The primary identification of the detected lines was done using the Lovas database of molecular lines detected in space~\cite{lovas09}\footnote{http://physics.nist.gov/restfreq}. However, some detected lines were absent from this database. Part of the absent lines have been found and identified by Kalenskii et al. \cite{kalen22}. The CDMS~\cite{muller01,muller05} and JPL~\cite{pickett} spectral line catalogs were used to identify the others. However, eight lines remained unidentified.

Practically all detected thermal lines have maxima at velocities near $-3$~km~s$^{-1}$. No lines with maxima near 9~km~s$^{-1}$ were detected. The lines of H$_2$CO at 72837.948 MHz and HC$_5$N at 85201.346 MHz are non-Gaussian and were approximated by two components each, with the radial velocities of the stronger components being near the system velocity and those of the weaker components being near $-1$~km~s$^{-1}$. It is possible that the remaining lines of these and some other molecules are also double lines, but the signal-to-noise ratio achieved does not allow us to separate them into components.

At 80578.283 MHz, we detected the $1_{1,0}-1_{1,1}$ HDO line, which has a radial velocity at maximum of 0.066~km~s$^{-1}$. However, the line identification cannot be considered reliable because only one line of this molecule has been detected. To verify the identification, it is best to observe the $2_{1,1}-2_{1,2}$ HDO line at 241561.550~MHz.

The radial velocity of the $N,J=2,2-1,1$~$^{34}$SO line is -0.92 km~s$^{-1}$, but the width of this line is very large (9.28 km~s$^{-1}$). It is probably blended with some unidentified line and the parameters given in the table are related to the whole blend.

Thus, it is not possible to separate the contributions of the different sources in our data. We can only conclude that most of the detected emission arises in the MM1 core and/or other objects with radial velocities close to the systemic velocity. In order to distinguish the contributions of individual cores, it is necessary to observe DR21OH with high spatial resolution.

A total of 69 molecules and their isotopic species were found at 4~mm, ranging from simple di- or triatomic molecules such as SO, SiO, and CCH, to complex organic compounds, CH$_3$CN, CH$_3$OCH$_3$ or CH$_3$CH$_2$OH. The list of detected molecules is given in %table~\ref{tab:detmol}; molecules found by spectral line stacking (see Section~\ref{sec:sls}) are shown in italics.
Table~1; molecules found by spectral line stacking (see Section~4) are shown in italics.

A significant fraction of the results are qualitatively similar to those of the 3 mm survey. The lists of molecules detected at 3 mm and 4 mm overlap to a large extent. The bulk is the species that are frequently observed in dense cores, such as HC$_3$N, SO, OCS, etc. However, a notable fraction of the detected molecules are characteristic of hot cores. In particular,  COMs such as CH$_3$OCHO, CH$_3$OCH$_3$, CH$_3$CH$_2$OH, CH$_3$COCH$_3$ are found, which, according to modern concepts, are formed in icy dust mantles and enter the gas phase during evaporation of mantles in hot cores or their sputtering by shock waves. However, it is well known that e.g., methanol, which is also formed in dust mantles, was also found in dense clouds and even dark clouds. All the lines of COMs we have detected (except several methanol lines) arise at transitions between levels with excitation energies less than 100~K. Therefore it is possible that these lines arise in dense cores and not in hot regions~(see Subsection~3.1 for further discussion of this issue). 

Apart from neutral molecules, various molecular ions -- HC$^{18}$O$^{+}$, HC$^{17}$O$^+$, H$^{13}$CO$^+$, DCO$^+$, etc. have been found. These ions, besides diffuse clouds, are often present in dense clouds with high optical absorption, where the primary sources of ionization are cosmic rays and X-ray radiation from protostars~(\cite[see][and references in this article]{padovani09}). 

Several deuterated compounds were found, including the deuterated cyanoacetylene DC$_3$N, which was only presumably found at 3 mm. The detection of two other molecules presumably found at 3 mm, CH$_2$NH and HCCN, was not confirmed.

\subsection{Rotation diagrams}
\label{subsec:rotdiagr}
Rotation diagrams (RDs) are a widely used method for estimating the temperature of interstellar gas and are described in a large number of papers \cite[e.g.][]{blake}. If several optically thin lines of the same molecule are found, then the column densities  at the upper energy levels of these lines, divided by the statistical weights of the levels ($N_u/g_u$), can be estimated. Under the Local Thermodynamic Equilibrium (LTE), the points on the plot of $\log (N_u/g_u)$ versus $E_u/k$, where $E_u$ is the upper level energy, $k$ is the Boltzmann constant, lie on a straight line, the slope of which is inversely proportional to the temperature. This temperature is called rotational temperature ($T_{rot}$). Since in molecular clouds LTE, if it exists, is achieved due to collisions with H$_2$ molecules, the rotational temperature often turns out to be close to the kinetic one. The point of intersection with the Y axis makes it possible to estimate the column density of the molecule.

\begin{figure}
\includegraphics[height=0.7\textheight]{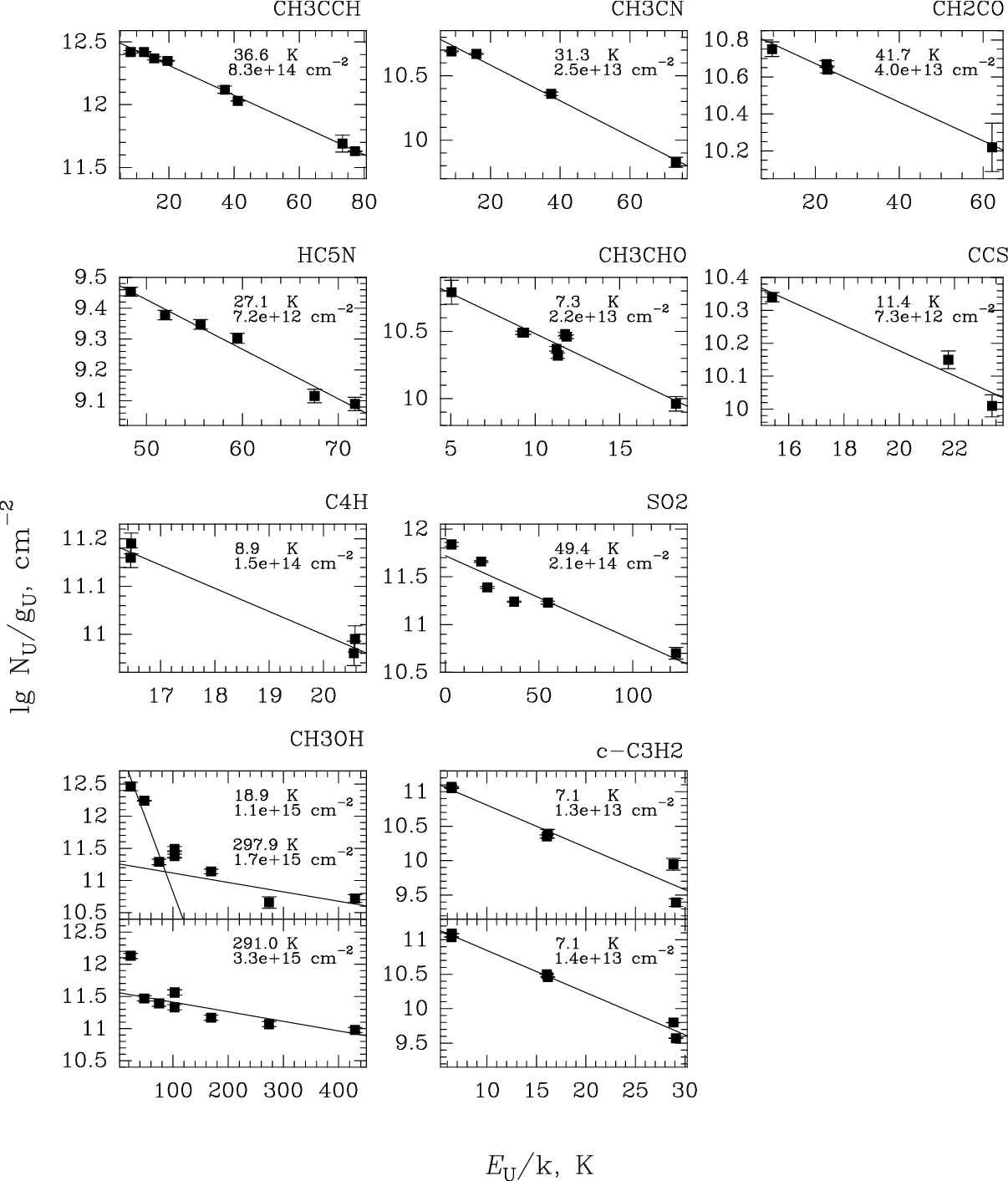}
\caption{Rotation diagrams plotted from lines found at 4~mm. RDs for methanol: upper panel -- plotted from the observed transitions; lower panel -- model RD plotted from the same transitions when they are optically thick. RDs for cyclopropenylidene: upper panel -- plotted from the observed transitions; lower panel -- plotted from the results of RADEX modeling ($T_{kin}=32$~K, N$_{\rm c-C3H2}=6.4\times 10^{12}$~cm$^{-2}$, $n_{\rm H2}=2\times 10^5$~cm$^{-3}$)}
\label{fig:rd1}
\end{figure}

We plotted an RD for each molecule with four or more isolated spectral lines detected\footnote{Excluding  molecules  whose individual spectral features are hyperfine components of the same rotational transition with practically the same upper-level energies}. These diagrams are shown in Fig.~\ref{fig:rd1}, and determined parameters are presented in Table~\ref{tab:rottab}. The RDs for four molecules: methyl acetylene (CH$_3$CCH), methyl cyanide (CH$_3$CN), cyanodiacetylene (HC$_5$N), and ketene (CH$_2$CO) demonstrate only slight point scattering around the approximating lines. The rotational temperatures fall within a narrow range, 27--42~K. This result is in a good agreement with the estimates of the kinetic temperature of MM2 (32~K), as well as of all other cores in DR21OH, except MM1, obtained by Mangum et al.~\cite{mangum92} from VLA observations of ammonia lines. Since symmetric-top molecules CH$_3$CCH and CH$_3$CN are good "thermometers" of the interstellar gas~\cite{askne,loren}, we believe that the temperature of the bulk of the gas in DR21OH is about 30--40~K.

The RD for methanol (CH$_3$OH) clearly shows the existence of two components. The first of them dominates the lines with $E_u/k < 100$~K (low-energy component), and the second dominates the lines with $E_u/k > 100$~K (high-energy component). The low-energy lines are approximated by a straight line corresponding to $T_{rot}=19$~K, and the high-energy lines,  by a straight line corresponding to $T_{rot}=298$~K. It is natural to interpret this diagram in such a way that low-energy lines appear in cold regions and high-energy lines, in hot regions. This interpretation is in agreement with the results of SMA observations of Minh~et~al.~\cite{minh12} and Orozco-Aguilera~et~al.~\cite{orozco19}. They found that methanol emission originates in hot cores MM1a and MM1b and in the lobes of a bipolar outflow south of MM1, elongated into the west-east direction for a distance of approximately $ 24''$, which at a distance to the source of 1.4 kpc corresponds to a distance projection on the picture plane of 0.15~pc. Low-energy lines that do not require high temperatures to arise are likely to occur in both of these regions; in addition, more extended emission regions, filtered out by the SMA array, may exist. High-energy lines may appear in a gas with a temperature on the order of 300~K, i.e., in hot cores discovered by Minh et al.~\cite{minh12}. However, this interpretation can be slightly modified. Figure~\ref{fig:rd1}, the bottom row shows a model RD constructed from the methanol lines we observed for the case when all their optical depths are high and therefore the brightness temperatures are the same. The sizes of the emitting region are chosen equal to the sizes of the hot core MM1a ($1.6''\times 1.1''$)~\cite{minh12}, the temperature of the region is chosen to be 100~K. Note that the resulting rotational temperature does not depend on the kinetic temperature of the source, but the position of the points along the Y axis depends on this parameter -- the higher the temperature, the higher the points are located. This RD strongly resembles the diagram plotted from actually observed high-energy lines, and the obtained value of $T_{rot}$ almost coincides with the rotational temperature of the observed high-energy component. Therefore, observed high-energy lines may be optically thick. In this case, the source temperature must be high because otherwise, it is problematic to obtain a large optical depth of lines with $E_u/k \sim 300-400$~K; however, it can differ greatly from the obtained value $T_{rot}=298$~K. 

The rotational temperatures for CH$_3$CHO, CCS, and c-C$_3$H$_2$ fall within the range 7--11~K. This can mean either that these molecules exist in a colder medium than CH$_3$CCH, CH$_3$CN, and HC$_5$N, or that the rotational level populations of these molecules are not thermalized by collisions. To deal with this issue, we modeled the c-C$_3$H$_2$ (cyclopropenylidene) emission for a number of source parameter sets using the Large Velocity Gradient (LVG) method with the RADEX code~\cite{radex}. The temperature was fixed at 32~K. The best agreement with the observations was obtained for the following parameter sets: n$_{\rm H2}=2\times 10^5$~cm$^3$ and N$_{\rm c-C3H2}=6.4\times 10^ {12}$~cm$^{-2}$. The rotation diagram for this set of parameters is shown in Fig.~\ref{fig:rd1} on the bottom right plot. It is very similar to the rotation diagram plotted from the observed c-C$_3$H$_2$ lines and gives the same rotational temperature (7 K). Therefore, we believe that the c-C$_3$H$_2$ emission arises at the same gas temperature as the CH$_3$CCH, CH$_3$CN, and HC$_5$N emission; the H$_2$ number density in the emitting region is about 2$\times 10^5$~cm$^{-3}$, and c-C$_3$H$_2$ column density is about $6.4\times 10^{12}$~cm$^{-2}$. %The model column density is approximately twice less than that obtained from the RD.

Unfortunately, it is impossible to carry out such modelling for ethanal (CH$_3$CHO) and thioethenylidene (CCS) because of the absence of collisional constants for these molecules. As for CH$_3$CHO, we {\em assume} that the low rotational temperature of this molecule is due to deviations of the level populations from LTE. The reason is as follows: ethanal is a prolate slightly asymmetric top. The same type of molecule is methanol, whose excitation was studied by Kalenskii \& Kurtz~\cite{kk16}. We believe that the rotation diagrams of two molecules of the same type may have the same regularities. Kalenskii and Kurtz~\cite{kk16} showed that the rotational temperature obtained from the $J_K-(J-1)_K$ methanol lines with the same $J$ and different $K$ values is lower than the kinetic temperature when H$_2$ number density is lower than $\lesssim 10^8$~cm$^{-3}$ and depends on the H$_2$ number density rather than on temperature. In our survey, only such lines of CH$_3$CHO\footnote{Except for the $2_{1,2}-1_{0,1}E$ line} were found, so it is natural to assume that just the set of lines caused the low rotational temperature.

%CCS  Образуется на ранних стадиях облаков, затем в центральных областях вымораживаются, непонятно, известно ли что-нибудь о химии на пылинках, затем освобождаются в горячих ядрах и за фронтами ударных волн.

Sulfur dioxide emission on the SMA maps of Minh~et~al.~\cite{minh12} is strongly concentrated toward the hot cores MM1b and MM1b. However, we obtained a rotational temperature of SO$_2$ equal to 48~K. Therefore one can assume that we observed SO$_2$ emission that arises in regions with moderate gas temperature. However, if the warm region radiating in the sulfur dioxide lines is compact, it is not clear why it is not visible in the maps of Minh et al. If it is extended, it is not clear why it is not revealed in the rotation diagrams of CH$_3$CN, CH$_3$CCH, and HC$_5$N. To deal with this question, we have modelled the emission of the SO$_2$ molecule with RADEX and found that, in a wide range of molecular gas parameters the rotational temperature of sulfur dioxide, determined from the transitions we observed, turned out to be much lower than the kinetic temperature. Therefore, our results do not indicate the existence of sources with gas temperatures on the order of 50 -- 60 K and do not contradict the assumption that sulfur dioxide emission arises in hot cores. However, while the maps of Minh et al~\cite{minh12} are dominated by emission from the MM1b core, the radial velocities of the lines detected by us are close to the systemic velocity that is characteristic of the MM1a core (see Introduction). It is natural to assume that the contribution of the MM1a core to these lines dominates. This difference in results may be due, for example, to the fact that Minh et al. observed higher energy transitions than those detected by us. While the energies of the upper levels of the lines observed by Minh et al. are of the order or higher than 200~K, similar values for the lines detected by us are of the order or lower than 120~K. If MM1a is colder than that of MM1b, it is possible that the emission of MM1a will be stronger in low-energy lines and that of the MM1b, in high-energy lines. Reliable measurements of the temperature of MM1a and MM1b are needed to test this assumption\footnote{The alternative assumption of the existence, in addition to MM1a, of some other hot region of gas with a velocity close to the systemic velocity seems unlikely for the same reasons as the assumption of the existence of a region with a temperature of about 50 K}.

\subsection{Column densities determined from single lines.}
\label{subsec:colden}
When it was not possible to construct a rotation diagram, we determined the column density of the molecule from single lines, assuming LTE. Exactly the same procedure was used as in the analysis of the survey of DR21OH at 3 mm~\cite{kalenskii10a}. Knowledge of the integrated intensity of a molecular line makes it possible to determine the column density at the upper level of the observed transition. Using this value, assuming LTE, and knowing the rotational temperature, one can calculate the total column density of the molecule. Rotational temperature cannot be determined from observations of a single line and must be chosen based on extraneous considerations. In this work, we assumed this temperature to be equal to the kinetic temperature of the gas in extended regions (32~K).

Many of the molecules that have been detected --- COMs CH$_3$OCHO, CH$_3$CH$_2$OH, CH$_3$OCH$_3$, etc. --- according to current concepts, are formed in the mantles of dust grains and rapidly pass into the gas phase as a result of mantle evaporation or sputtering in shock-heated gas. However, such molecules are observed not only in hot cores but, albeit with less abundances, in dense cores and even dark clouds (e.g.~\cite{agundez23,spezzano}). In observations with low spatial resolution, the contribution of cold regions may dominate, as usually they are much more extended than hot cores. From our data, taken in one direction, it is impossible to determine where the emission of the detected COMs arises in compact hot cores or in extended dense clouds. In the case of methanol, which is also a COM that forms on dust grains, we see two components: a cold and a hot~(see Subsection 3.1). In low-energy lines, the contribution of the cold component predominates. The lines of all other COMs found at 4~mm are low-energy, so it is natural to assume that they originate mainly in cold gas. Therefore we took the rotational temperature equal to 32 K in all cases.

For a number of molecules, the detected lines are hyperfine components of the same rotational transition. It is impossible to construct a rotation diagram from such lines because the energies of their upper levels practically coincide. For many molecules -- H$_2$CO, HC$_3$N, OCS, etc. only two spectral lines were found. In addition, for C$_4$H, CCS, and CH$_3$CHO, low rotational temperatures, obtained from the rotational diagrams, indicate strong deviations from LTE, which could distort the column densities. In each of these cases, we determined the column density for each line, assuming $T_{rot}=32$~K, and found the average value.

Column densities determined from single lines are given in Table~\ref{tab:colden}. The last 11 lines show the column densities of molecules determined by spectral-line stacking (see Section ~\ref{sec:sls}). The third column shows the relative abundance of molecules. The H$_2$ column density was determined from the integrated intensity of the $1-0$ C$^{18}$O line, taken from the results of the DR21OH spectral survey at 3~mm~\cite{kalenskii10a}. The abundance of C$^{18}$O was taken equal to $10^{-7}$, which approximately corresponds to the average value of the abundance of C$^{18}$O in dense cores (see Fig.~18 in~\cite{baras21}). With this abundance of C$^{18}$O, the H$_2$ column density turned out to be equal to $3\times 10^{23}$~cm$^{-2}$.

The obtained abundances are given in Tables~\ref{tab:rottab} and \ref{tab:colden}. For seven molecules (H$^{13}$CN, H$^{13}$CO$^+$, HC$^{18}$O$^+$, SO, $^{34}$SO, HCS$^+$, OCS), these abundances can be compared with the values obtained for a large number of dense cores by Rodriguez-Baras et al.~\cite{baras21}. Their paper shows the dependencies of molecular abundances on various cloud parameters, including density. For all seven molecules, a negative correlation between the density and abundance was found in~\cite{baras21}. The abundances obtained by us correspond to density intervals from $\sim 10^4$ to $\sim 10^5$~cm$^{-3}$. This is lower than the density obtained by us from CCS lines (see Subsection ~\ref{subsec:rotdiagr}) and by 2 orders of magnitude lower than the densities obtained from dust emission by Mangum et al.~\cite{mangum91} in MM1 and MM2. This discrepancy may be a consequence of the structure of the source. Mangum et al. estimated the density in clumps MM1 and MM2. It is possible that the CCS emission appears at the periphery of the clumps, where the density is lower, while the emission of the seven listed molecules occurs in an even more rarefied interclump medium. However, the fact that the abundances of all seven molecules correspond to approximately the same density, which is lower than that obtained from the CCS lines, makes the following explanation more likely. Figure~18 in the paper~\cite{baras21} shows that the interval of C$^{18}$O abundances in dense cores is about an order of magnitude. Probably, in DR21OH-M C$^{18}$O abundance is close to the lower boundary of this interval, i.e., approximately three times lower than the average value. In this case, the abundance of molecules in Table~\ref{tab:rottab} and \ref{tab:colden} is overestimated by a factor of three. The results of Rodriguez-Baras et al.~\cite{baras21} show that in this case the actual abundances of H$^{13}$CN,  H$^{13}$CO$^+$, HC$^{18}$O$^+$, SO, $^{34}$SO, HCS$^+$ and OCS in DR21OH correspond to a density range of approximately $3\times 10^4-3\times 10^5$~cm$^{-3}$, which does not contradict the density obtained from the CCS lines.

\section{Spectral line stacking}
\label{sec:sls}
If the abundance of a molecule in a molecular cloud is below $\sim 10^{-11}-10^{-12}$, then, as a rule, with the state-of-art equipment it is impossible to detect the radio lines of this molecule in a reasonable observing time. However, often the number of lines of the molecule falling within the observed frequency range is very large, reaching hundreds or thousands. This is especially true for complex molecules. In this case, it is possible to stack these lines and for the aggregate spectral detail to obtain a signal-to-noise ratio higher than that achieved for the strongest lines of the sought-for molecule. Therefore, spectral line stacking (SLS) makes it possible to find even the molecules whose lines are not visible. This method was first applied to search for molecules by Johansson et al.~\cite{johansson}. Kalenskii and Johansson~\cite{kalenskii10a,kalenskii10b} used SLS to analyze the results of surveys of star-forming regions DR21OH and W51e1/e2 in the 3-mm wavelength range and expanded the lists of detected molecules\footnote{Johansson et al. and Kalensky and Johansson called this method "composite averages"}. A 13-atom molecule benzonitrile (C$_6$H$_5$CN) and several other, even more complex molecules, have been discovered in space for the first time using SLS~\cite{kalen17,mcguire18,loomis}.

The method is described in a number of papers~\cite{johansson,kalenskii10a,loomis}; however, the technique used here differs in detail from those used in these works. Therefore in this section, we present our approach.

When searching for a molecule using SLS, the Lines of Other Molecules (LOMs) are interferences that can lead to a spurious "detection" of the sought-for molecule. This can occur due to random coincidences in  frequency between LOMs and the lines of the sought-for molecule. The impact of LOMs had to be minimised. For this purpose, we excluded from further consideration the parts of the spectrum occupied by LOMs. This was done as follows: first, the model spectrum (MS) of all already detected molecules in DR21OH was built. For this purpose, we used the WEEDS extension of the CLASS software, which, in turn, assumes LTE. The source temperature was assumed to be 32~K. The frequencies and other necessary parameters of the LOMs were taken from the spectral line catalogs CDMS~\cite{muller01,muller05} and JPL~\cite{pickett}. The column densities for calculating line intensities were taken from Tables~II and~III. The widths of all lines were taken to be 4.24 km s$^{-1}$, which is the mean FWHM in DR21OH according to our data. 

This MS makes it possible to exclude all frequency ranges occupied by the emission of already detected molecules. For this purpose, we marked every frequency interval where the MS brightness temperature exceeds the noise rms of the observed spectrum {\em provided that the maximum MS brightness temperature within this interval exceeds 2~rms}. Then the width of each such interval was increased by 20\% to account for the possible differences in line widths. The spectral lines of the sought-for molecule that fell into these intervals were excluded from consideration. In addition, the observed emission within these frequency intervals was replaced by noise with the same rms as the noise rms of the observed spectrum. This is necessary for the correct assessment of the weights with which the lines are stacked (see Appendix).

Next, narrow sections of the same width (200 channels, which corresponds to a bandwidth of 61 MHz), centered on the frequencies of the sought-for molecule lines, were cut out from so prepared broadband source spectrum and stacked with weights chosen in a certain way (the construction of a stacked spectrum is described in more detail in the Appendix). The weights depend on the rotational temperature of the sought-for molecule, which is unknown in advance. Therefore at the first stage, the stacked spectra were built for seven temperatures: 9.375, 18.75, 37.5, 75, 150, 225, and 300 K. These temperatures were chosen because just for them JPL and CDMS catalogs provide molecular partition functions\footnote{Partition functions of some molecules for 2.725, 5.0, 500, and 1000~K are additionally given in the Cologne catalog, but we didn't build stacked spectra for these temperatures.}.

\begin{figure}
    \centering
    \includegraphics[height=0.5\textheight]{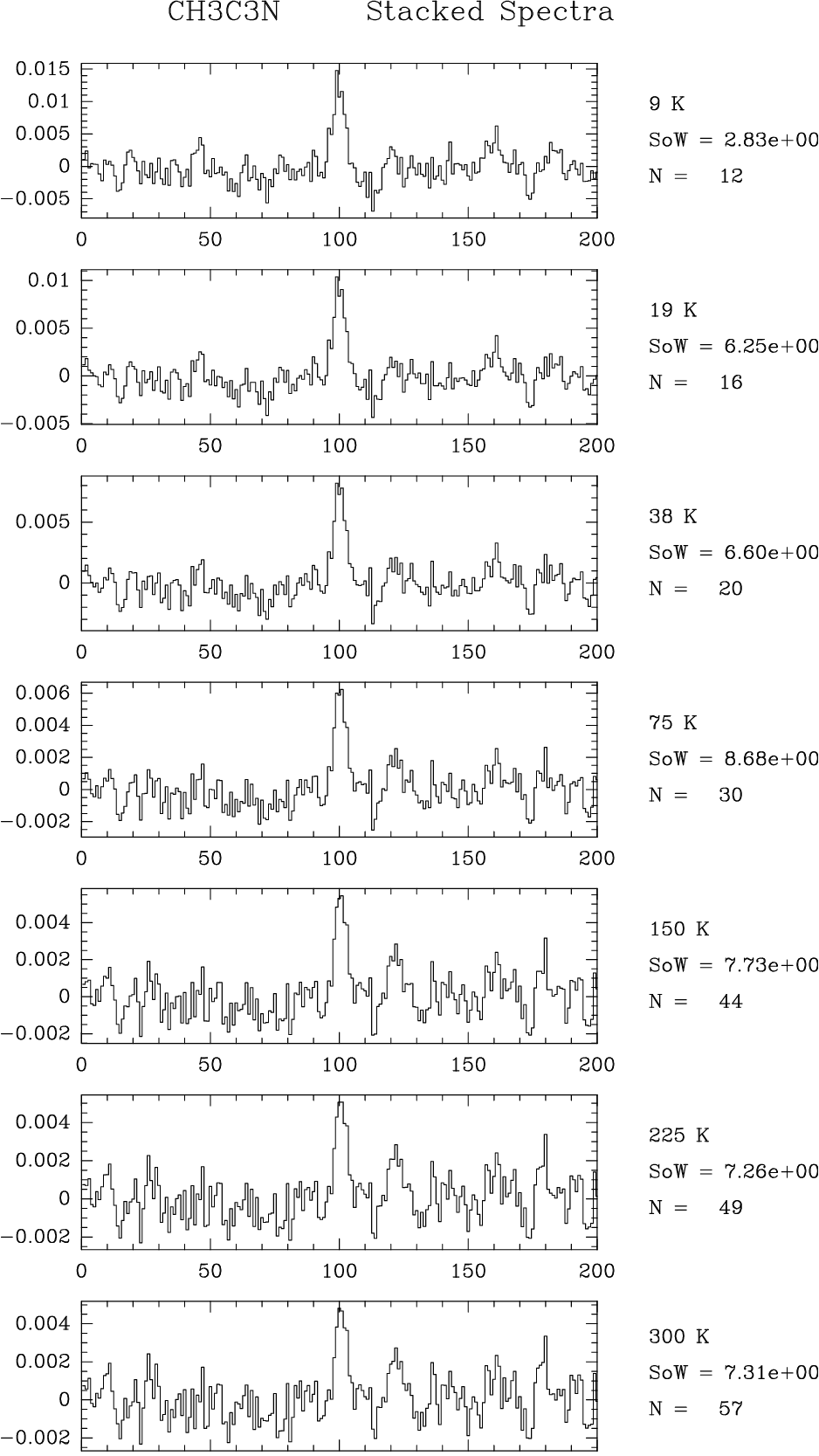}
    \caption{Stacked spectra of methylcyanoacetylene (CH$_3$C$_3$N) for seven rotation temperatures.  N -- the number of stacked lines, SoW -- the sum of weights of stacked lines.}
    \label{fig:ch3c3n}
\end{figure}

As a result, seven stacked spectra are obtained for the sought-for molecule~(see Fig.~\ref{fig:ch3c3n}). If the molecular emission is present in the observed spectrum, a spectral feature appears in the center of the stacked spectrum, which we call a stacked line (SL). The signal-to-noise ratio for SL in the stacked spectrum is higher than this ratio even for the strongest lines of the sought-for molecule in the observed spectrum.

We performed this procedure for all molecules whose spectral line data are available in the JPL and CDMS databases. As a result, several dozens of stacked lines were detected. However, for each source and frequency range, there is a certain confusion level, upon reaching which the observer is faced with a situation when weak lines of different molecules can have practically the same frequency. When observing at the millimeter wavelengths using the state-of-art equipment, it is quite possible to reach the confusion level or come close to it. This can lead to spurious detections, an example of which is the "discovery" of glycine~\cite{kuan}, subsequently criticized by Snyder et al.~\cite{snyder}. Therefore, each detected SL should be additionally tested. We applied two tests for this purpose. The first of them is as follows: one can build a model stacked spectrum (MSS) using MS instead of the observed spectrum. If the detection is spurious and SL is a result of a random superposition of weak lines of already detected molecules, then it must appear on the MSS as well. The absence of SL on the MSS is an argument in favor of the statement that the detection of the sought-for molecule is real.
 
Snyder et al.~\cite{snyder} proposed criteria for a molecule to be reliably detected, which can be summarized as follows: (1) many lines of this molecule must be detected; (2) the ratio of line intensities must correspond to some rotational temperature depending on the source parameters. In particular, the lines that,  according to their spectral characteristics, {\em should be} reliably detected, {\em must be} detected. The method of spectral line stacking fundamentally does not allow the application of these criteria. Instead, we used the fact that line weights are chosen based on line parameters and rotational temperature and repeatedly applied the following procedure as a second test: for the molecule under test, we did stacked spectra with randomly selected weights. If the SL on the "correct" stacked spectrum appears due to the emission of the sought-for molecule, stacking with random weights will  weaken the SL or lead to its disappearance. But if the SL appears due to a random coincidence in the frequency of weak lines of other molecules, such stacking can both weaken and strengthen the SL. 

\begin{figure}
    \centering
    \includegraphics[height=0.3\textheight]{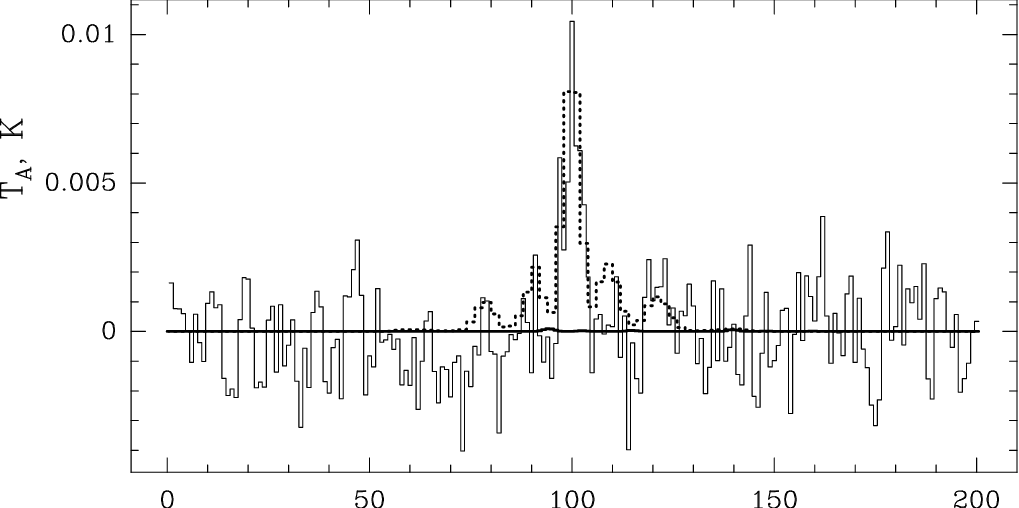}
    \caption{The thin solid line is the synthetic spectrum of methylcyanoacetylene (CH$_3$C$_3$N), constructed from observational data. The thick solid line is a similar synthetic spectrum built using the model spectrum of the source (MSS), which includes all detected molecules, except for methylcyanoacetylene. The dotted line is a synthetic spectrum constructed from a spectral model that includes only methylcyanoacetylene.
    }
    \label{fig:ch3c3nfin}
\end{figure}

If an SL successfully passes both tests, the molecule is considered to be detected. Its column density was estimated, and its emission was added to the model spectrum. The column density was estimated as follows: first, some a priori column density $N^a$ was assigned to this molecule, then a model spectrum of only this molecule was constructed (specific model spectrum), and the stacked spectrum was built from this model spectrum. Then the molecule column density $N_{mol}$ can be found from the obvious equation:  
\begin{equation}
N_{mol}=\frac{T_R^{sl}}{T_R^{sm}} \cdot N^a
\end{equation}
where $T_R^{sl}$ is the SL brightness temperature and $T_R^{sm}$ is the brightness temperature of the stacked line, constructed from the specific model spectrum. An example of the comparison of the "observed" and "specific" SLs is presented in Fig.~\ref{fig:ch3c3nfin}.

The column densities of two molecules found by the traditional method, ethyl alcohol (CH$_3$CH$_2$OH) and dimethyl ether (CH$_3$OCH$_3$), determined from single lines are extremely unreliable. The reasons are as follows. As concerns ethyl alcohol, only one weak line with poorly determined parameters was found. As for dimethyl ether,  all its lines are grouped into blends, and it is impossible to determine the parameters of single lines. Therefore, we determined the column densities of these molecules in the same way as the column densities of molecules found by spectral line stacking. The results are shown in Table~\ref{tab:colden}.

Nine molecules (see Table~1) passed both tests. However, the SL integrated intensity of three of them is small (less than its five errors), so the detection of these molecules is still questionable and should be further tested.

%CH$_3$CCCN - обн. $N_{CH3CCCN}$=4E+11! Почему так мало?
%Не учтена Beff.
%Кайзер: CN•+CH3CCH -> CH3CCCN+H•;
%Ищем лучевую конц. для 32К.

\section{Summary}
The results of a survey of the region of massive star formation DR21OH in the 4 mm wavelength range are presented. Sixty-nine molecules and their isotopologues have been detected, ranging from simple diatomic or triatomic molecules, such as SO, SiO, CCH, to complex species such as CH$_3$OCHO or CH$_3$OCH$_3$. A notable part of the results qualitatively repeats the results of the survey of the same source at 3 mm. The lists of molecules found at 3~mm and 4~mm are broadly overlapped. The bulk of the molecules detected at 4~mm are commonly found in dense cores, e.g., HC$_3$N or CH$_3$CCH. Rotation diagrams constructed from CH$_3$CN, CH$_3$CCH, and HC$_5$N lines showed that the gas temperature in DR21OH is on the order of 30--40~K.

Eighteen of the detected species are COMs. According to modern concepts, these molecules are formed in the mantles of interstellar dust grains and appear in the gas phase either as a result of mantle evaporation in hot cores around protostars or due to nonthermal desorption in colder gas. Unfortunately, from the data of our single-position survey, it is impossible to unambiguously establish in which regions the observed emission of COMs arises, in hot cores or in the surrounding gas, like the emission of the other detected molecules. Only for methanol (CH$_3$OH) we found a number of high-energy transitions and separated two components, cold and hot. The cold component produces low-energy lines with level energies below 100~K, while the hot component produces high-energy lines with level energies above 100~K. Low-energy lines can appear in the same regions as the CH$_3$CCH, CH$_3$CN, HC$_5$N lines. High-energy lines can appear in gas with temperatures above 100~K, e.g., in hot cores discovered by Minh et al.~\cite{minh12}. All other COMs demonstrate only low-energy lines, which {\em presumably} originate in colder gas.

The column densities and the abundances of the detected molecules were estimated. For a number of simple molecules, the abundances agree with those that are typical for dense clouds at an H$_2$ number densities of the order of $10^4-10^5$~cm$^{-3}$.

Nine molecules, including COMs CH$_3$C$_3$N, CH$_3$CH$_2$CN, CH$_3$COCH$_3$, were found using spectral-line stacking. This fact demonstrates that spectral line stacking may be a useful tool for the analysis of broadband spectra.

\begin{acknowledgments}
SVK was financially supported through the Ministry of Science and Higher Education of the Russian Federation grant No. 075-15-2021-597, and EAM, through the Russian State program GZ019ASC009 109.61. The authors are grateful to P. Bergman and A. O. H Olofsson (Onsala Space Observatory) for help with the application and observations and to an anonymous referee for helpful comments. The Onsala Space Observatory national research infrastructure is funded through Swedish Research Council grant No. 2017-00648. This paper makes use of NASA’s Astrophysical Data System as well as the following data sets: Observed Interstellar Molecular Microwave Transitions~\citep{lovas09}, the Cologne Database for Molecular Spectroscopy~\citep{muller01,muller05}, the JPL Spectral Line Catalog~\citep{pickett}; 
\end{acknowledgments}

\appendix

\section{Line stacking procedure}
This Appendix describes the process of spectral-line stacking itself when the sought-for molecule and the rotational temperature have already been chosen. LTE is assumed elsewhere in this section. First, from the JPL or CDMS catalog we selected the lines of the sought-for molecule, whose frequencies fall within the range of the survey. From the obtained inventory we excluded the lines whose frequency errors and/or upper-level energies exceed some threshold values. In this work, the threshold for the frequency error was set equal to 0.5 MHz. Near the low-frequency boundary of the survey band, this value corresponds to 2.1 km s$^{-1}$, i.e., to the mean half-width at half maximum of the lines in DR21OH. The threshold for the upper-level energy was set equal to 1000~K.

Further, using the line parameters from the catalogs, we searched for the strongest line at the chosen rotational temperature (hereinafter referred to as the reference line). When the column density of the molecule is unknown, we cannot find the  brightness temperatures of molecular lines, but assuming LTE we can find the brightness temperature ratios. This is enough to correctly determine the line weights.

Intensity $I$ (cm$^{-1}$/(molecule/cm$^2$)), presented in the JPL and CDMS catalogs for each molecular line is proportional to the maximum optical depth of the line $\tau^{max}$:
\begin{equation}
I=\frac{\kappa_\nu}{f(\nu)\cdot n} = \frac{\tau(\nu)}{f(\nu)\cdot N} \propto \frac{\tau^{ max}\Delta\nu}{N}
\end{equation}
where $N$ is the column density of the molecule, $\tau(\nu)$ is the optical depth of the line as a function of frequency, $f(\nu)$ is the normalized function describing the line profile, $\Delta\nu$ is the line width (FWHM). From here, one can find the maximum optical depth of the line (up to a certain coefficient, which depends on $N$ and is the same for all lines). Further, among the selected lines, the strongest one is found, that is, the line with the largest  optical depth. Some optical depth value is {\em assigned} to this line. Then, the optical depths of the remaining lines can be found using the relation:
\begin{equation}
\tau_i^{max}=\tau_{ref}^{max}\times I_i/I_{ref}\times \frac{\nu_i}{\nu_{ref}}
\end{equation}
It is assumed here that the widths of all lines in velocity units are the same, in which case $\Delta\nu_i \propto \nu_i$. Knowing the optical depth $\tau_i$, one can find the amplitude of each line, $\Delta T_i$, using the well-known relation:
\begin{equation}
\Delta T_i=(T_{ex}-T_{bg})\times (1-exp(-\tau_i))
\end{equation}
where $T_{ex}$ and $T_{bg}$ are the Rayleigh--Jeans equivalent temperatures of excitation and microwave background. The excitation temperature under LTE is equal to the chosen rotational temperature.

When the brightness temperature of each line is known, one can calculate the ratios $R_i=T_i/T_{ref}$, where $T_{ref}$ is the brightness temperature of the reference line and $T_i$ is the brightness temperature of the $i-$th line. All lines with $R_i \lesssim 0.01$ are excluded from further consideration. Then, from the observed spectrum we cut out narrow bands centered on the frequencies of selected lines and divided the brightness temperatures in each band by the relevant $R_i$. After this procedure, the brightness temperatures of the lines in all narrow-band spectra became approximately the same, and the noise rms increased proportionally to $R_i$. The narrow-band spectra were stacked with weights inversely proportional to the spectrum noise dispersion and the weight of the strongest line equal to unity, and the results were divided by the sum of the weights. With this method, the SL amplitude turns out to be approximately equal to the reference line amplitude, and the noise level is less than that in the original spectrum by $\sqrt{\sum (w_i)}$ times.

The sizes of the sources in which we are looking for new molecules are unknown. If we assume that these sizes are comparable to the main beam width, then the weak lines that appear in such sources should be optically thin. The optical depth of the strongest line $\tau_{ref}^{max}$ is set equal to 0.1 by default, and in this survey, we keep this value of the maximum optical depth. The exact value of the optical depth provided that $\tau_{ref}^{max} << 1$, does not matter for calculating the weights. However, compact sources may fall into the telescope beam. For example, in our observations, it was hot cores MM1a and MM1b. The sizes of these sources are much smaller than the beam width, and even weak lines arising in these cores may be optically thick. In this case, the choice of line weights, made on the assumption that the lines are optically thin, is no longer optimal. 
Therefore, at the final stage, one should repeat SLS abandoning the assumption that the lines are optically thin. We rebuilt all stacked spectra, setting the maximum optical depth of the reference line of each molecule to be equal to 10. However, no new molecule was found as a result of this procedure.

\newpage

\clearpage\newpage

\clearpage\newpage

\newpage

\clearpage\newpage
\begin{table}
\label{tab:detmol}
\caption{Molecules found in the DR21OH star-forming region}
\begin{tabular}{|l|p{0.65\textwidth}|}
\noalign{\medskip}
\hline%\noalign{\smallskip}
Diatomic \rule{0pt}{7pt}&
NS, SiO, $^{29}$SiO, $^{30}$SiO, $^{30}$Si$^{18}$O?, $^{34}$SO, SO, SO$^+$\\
% 8
\hline
Triatomic \rule{0pt}{7pt}&
CCH, $^{13}$CCH, C$^{13}$CH, CCD, CCS, DCO$^+$, DCN, DNC, HCO, HDO, HNO, HC$^{18}$O$^+$, HC$^{17}$O$^+$, HCS$^+$, H$^{13}$CN, HC$^{15}$N, H$^{13}$CO$^+$,N$_2$O, N$_2$D$^+$, OCS, O$^{13}$CS?, {\em OC$^{34}$S?}, SO$_2$ \\
% 23
\hline
Four-atom \rule{0pt}{7pt}&
l-C$_3$H, {\em C$_3$S?}, H$_2$CO, H$_2^{13}$CO, H$_2$CS, HOCO$^+$, HOCN?, D$_2$CS, NH$_2$D\\
% 9
\hline
Five-atom\rule{0pt}{7pt}&
c-C$_3$H$_2$, l-C$_3$H$_2$, C$_4$H, CH$_2$CO, CH$_2$CN, DC$_3$N, H$^{13}$CCCN, HC$^{13}$CCN, HCC$^{13}$CN, t-HCOOH, HC$_3$N\\
% 11
\hline
Six-atom\rule{0pt}{7pt}&
CH$_3$CN, CH$_3$OH, {\em $^{13}$CH$_3$OH,} CH$_3$SH, NH$_2$CHO\\
% 5
\hline
Seven-atom\rule{0pt}{7pt}&
CH$_3$CCH, CH$_3$CHO, CH$_2$CHCN, {\em CH$_3$NH$_2$}, HC$_5$N\\ 
% 5
\hline
Eight-atom\rule{0pt}{7pt}&
CH$_3$OCHO, {\em C$_2$H$_3$CHO,} {\em CH$_3$C$_3$N,}\\
% 3
\hline
Nine-atom \rule{0pt}{7pt}&
C$_2$H$_5$OH, CH$_3$OCH$_3$, {\em CH$_3$CH$_2$CN,} {\em CH$_3$C$_4$H?}\\
% 4
\hline
Ten-atom\rule{0pt}{7pt}&
{\em CH$_3$COCH$_3$}\\
% 1
\hline
\end{tabular}
\flushleft
\scriptsize\bf The names of molecules detected by spectral line stacking are in italics. The question mark indicates molecules whose detection cannot be considered reliable.
\end{table}

\clearpage\newpage
\begin{table}
\caption{Rotational temperatures, column densities, and abundances of molecules obtained using rotation diagrams. Numbers in brackets show $1\sigma$ errors.
\label{tab:rottab}}
\begin{tabular}{|l|c|l|r|}
%\tablewidth{0pt}
\hline
Molecule       & $T_{rot}$  & $N_{mol}$ & $N_{mol}/N_{H_2}$$^1$\\
               &  (K)       &($10^{13}$~cm$^{-2}$)
                                        & ($10^{-10}$)\\
\hline
C$_4$H         & 8.9(0.8)	& 15(3.0)	&5(1)\\
c-C$_3$H$_2$   & 7.1(0.9)	& 1.3(0.53)	&0.4(0.2)\\
CCS            & 11.4(2.0)	& 0.73(0.28)&0.2(0.1)\\
CH$_3$CCH      & 35.6(3.9)	& 66(9.4)   &22(3)\\
HC$_5$N        & 28.7(1.6)	& 0.68(0.08)&0.2(0.02)\\
CH$_3$CN       & 31.3(1.6)	& 2.5(0.18) &0.8(0.06)\\
CH$_3$CHO      & 9.7(2.2)	& 2.5(0.86) &0.8(0.3)\\
CH$_3$OH       & 19(4)	    & 110(100)  &37(33)\\
CH$_3$OH$^2$   & 298(196)   & 170(170)  & -- \\
CH$_2$CO       & 46.1(3.8)	& 4.5(0.35) &1.5(0.1)\\
SO$_2$         & 49.4(7.4)	& 21(4.1)   & -- \\
\hline
\end{tabular}

\medskip
$^1-N_{H_2} = 3\times 10^{23}$~cm$^{-2}$ (see Subsection~3.1).\\
$^2$--abundance is not given because the size of the hot region and the column density of hydrogen in it are unknown.
\end{table}

\clearpage\newpage
\begin{table}
\caption{Column densities of molecules determined from single lines
\label{tab:colden}}
\begin{tabular}{|l|c|c|}
%\tablewidth{0pt}
\hline
Molecule & Column density        & $N_{mol}/N_{H_2}$\\
         & ($10^{12}$~cm$^{-2}$) &($ 10^{-11}$)\\
\hline
$^{13}$CCH	&	34	&	10.5	\\
C$^{13}$CH	&	47	&	14.5	\\
CCH	        &  2133	&   660	\\
C$_4$H		&	94	&	29.1\\	
CCD	        &	118	&	36.5	\\	
CCS	        &	10	&	3.1	\\	
CH$_2$CHCN	&	3	&	1.0	\\	
CH$_2$CN	&	1	&	0.3	\\
CH$_3$CHO	& 115.8 &   35.8 \\	
CH$_3$OCH$_3$&	110	&	34.1\\	
CH$_3$OCHO	& 100.3	&   31.1 \\
CH$_3$OH	&	3359.5	&	1040.1\\	
D$_2$CS		&	1.9	&	0.6	\\	
DC$_3$N		&	0.7	&	0.2	\\	
DCN	        &	24.1	&	7.5\\	
DCO$^+$		&	3.4	&	1.1	\\	
DNC	        &	8.4	&	2.6	\\	
H$^{13}$CCCN	&	1.1	&	0.4\\	
H$^{13}$CN	&	43.9	&	13.6\\	%
H$^{13}$CO$^+$	&	18.3	&	5.6	\\	
l-C$_3$H$_2$	&	0.9	&	0.3	\\	
H$_2$CO		&	582.2	&	180.2\\	
H$_2$CS		&	136.6	&	42.3\\	
H$_2^{13}$CO	&	23.8	&7.4\\	
HC$^{13}$CCN	&	1.5	&	0.5	\\	
HC$^{15}$N	    &	9.5	&	2.9\\	
HC$^{18}$O$^+$	&	2.7	&	0.8\\	
HCC$^{13}$CN	&	1.3	&	0.4	\\	
HC$_3$N		&	58.4	&	21.9\\	
\hline
\end{tabular}
\end{table}

\addtocounter{table}{-1}
\clearpage\newpage
\begin{table}
\caption{(continued)
%\label{tab:colden}
}
\begin{tabular}{|l|c|c|}
%\tablewidth{0pt}
\hline
Molecule & Column density        & $N_{mol}/N_{H_2}$\\
         & ($10^{12}$~cm$^{-2}$) &($10^{-11}$)\\
\hline
HCO	        &   83.2    &   25.8\\	
t-HCOOH         &   24.2    &   7.5\\
HCS$^+$		&	11.5	&	3.6	\\	
HDO		&	10.1	&	3.1	\\	
HN$^{13}$C	&	15.9	&	4.9	\\	
HNO		&	16.7	&	5.2\\	
N$_2$D$^+$	&	1.1	&	0.4\\	
NH$_2$CHO	&	1.1	&	0.4	\\	
NS		&	91.6	&	28.4\\	
OCS		&	101.7	&	31.5\\	
$^{29}$SiO	&	1.5	&	0.5\\	
%$^{30}$SiO	&	1.2	&	0.4	\\	
$^{34}$SO	&	25	&	7.8\\	
SO		&	405.4	&	125.5\\	
SO$^+$		&	36.8	&	11.4	\\	
%SO$_2$		&	174.4	&	51.0\\	
SiO		&	24.2	&	7.5\\	
%c-C$_3$H$_2$	&	55.6	&17.2\\	
l-C$_3$H	&	7.9	&	2.5\\	
l-C$_3$H$_2$	&	0.9	&	0.3\\	
%t-CH$_3$CH$_2$OH &	36$^1$	&	11.1\\	
\hline						
CH$_3$C$_3$N	&	1.1	&	0.4\\	
CH$_3$CH$_2$CN	&	1.2	&	0.4\\	
CH$_3$CH$_2$OH	&	36	&	11.1\\	
C$_3$S		&	0.3	&	0.1\\	
$^{13}$CH$_3$OH	&	59	&	18.3\\	
CH$_3$C$_4$H	&	4.5	&	1.4\\	
CH$_3$OCH$_3$	&	125	&	38.7\\	
CH$_3$SH	&	36	&	11.1\\	
CH$_2$CN	&	1.4	&	0.4\\	
OC$^{34}$S	&	4.9	&	1.5\\	
C$_2$H$_3$CHO	&	1.8	&	0.5\\
\hline
\end{tabular}
\end{table}

\clearpage\newpage
\begin{table}
\caption{Spectral lines of the DR21OH M star formation region in the 4-mm wavelength range.
\label{tab:detlines}}
\medskip
\begin{tabular}{|l|l|l|llll|l|}
%\tablewidth{0pt}
\hline
Frequency & Molecule & Transition & $\int T_RdV$ & $V_{LSR}$ & $\Delta V$ & $T_R$ & Notes\\
 (MHz)  &          &        &(K km s$^{-1})$&(km s$^{-1}$)&(km s$^{-1}$ )& (K)   & \\
\hline
68305.68	&CH$_3$OH		&$1_{1,0}-2_{0,2} E$		&0.27(0.04)	&-2.92(0.61)	&9.40(1.63)	&0.0265	&\\ 
68354.502	&CH$_3$CCH		&$4_3-3_3	$		&0.13(0.02)	&-3.33(0.05)	&3.98(0.00)	&0.0314	&bl\\ 
68361.035	&CH$_3$CCH		&$4_2-3_2	$		&0.30(0.02)	&-3.33(0.05)	&3.98(0.00)	&0.0714	&bl\\ 
68364.955	&CH$_3$CCH		&$4_1-3_1 	$		&0.67(0.02)	&-3.33(0.05)	&3.98(0.00)	&0.1586	&\\ 
68366.262 	&CH$_3$CCH		&$4_0-3_0	$		&0.81(0.02)	&-3.33(0.05)	&3.98(0.06)	&0.1901	&\\ 
68699.381       &H$_2$CS                &$2_{0,2}-1_{0,1}$              &0.69(0.02)     &-3.08(0.07)    &4.27(0.17)     &0.152 &\\ 
68972.154 	&SO$_2$			&$6_{1,5}-6_{0,6}$		&0.81(0.02)	&-3.03(0.09)	&6.00(0.22)	&0.1275	&rw\\ 
69002.890	&NS			&$J=3/2-1/2~F=5/2-3/2 e$	&0.45(0.02)	&-2.76(0.07)	&3.73(0.08)	&0.1123	&\\ 
69017.895	&NS			&$J=3/2-1/2~F=3/2-3/2 e$	&0.18(0.02)	&-2.76(0.07)	&3.73(0.00)	&0.0452	&\\ 
69037.336	&NS			&$J=3/2-1/2~F=3/2-3/2 e$	&0.17(0.02)	&-2.76(0.07)	&3.73(0.00)	&0.0415	&\\ 
69040.324	&NS			&$J=3/2-1/2~F=1/2-1/2 e$ 	&0.11(0.01)	&-2.76(0.07)	&3.73(0.00)	&0.0279	&\\ 
69227.183       &HC$_5$N                &$26-25$                        &0.31(0.02)     &-2.73(0.14)    &4.56(0.27)     &0.065  &\\
69281.115       &CCS                    &$N_J=5_6-4_5$                  &0.33(0.02)     &-3.23(0.10)    &3.99(0.24)     &0.078  &\\
69330.592       &NS                     &$J=3/2-1/2~F=3/2-3/2~f$        &0.14(0.01)     &-3.38(0.19)    &3.77(0.38)     &0.035  &\\
69408.371	&SO$^+$			&$^2\Pi_{1/2}~J=3/2-1/2e$	&0.17(0.02)	&-2.92(0.20)	&4.24(0.87)	&0.0380	&\\ 
69411.943	&NS			&$J=3/2-1/2~F=5/2-3/2 f$	&0.46(0.02)	&-3.36(1.31)	&3.71(0.31)	&0.1158	&\\ 
69437.850       &NS                     &$J=3/2-1/2~F=1/2-1/2~f$        &0.14(0.02)     &-3.33(0.16)    &3.22(0.42)     &0.040  &\\
69485.223       &NS                     &$J=3/2-1/2~F=3/2-1/2~f$        &0.17(0.01)     &-3.23(0.11)    &2.99(0.29)     &0.052  &\\
69575.923	&SO$_2$			&$1_{1,1}-0_{0,0}$	        &0.42(0.02)	&-2.99(0.57)	&5.20(0.05)	&0.0752	&\\ 
69746.720       &H$_2$CS                &$2_{1,1}-1_{1,0}$              &0.91(0.01)     &-3.19(0.03)    &4.10(0.07)     &0.209  &\\
69783.846       &SO$^+$                 &$^2\Pi_{1/2}~J=3/2-1/2f$       &0.11(0.01)     &-3.17(0.11)    &2.55(0.26)     &4.195  &\\
70534.033 	&H$^{13}$CCCN		&$8-7$				&0.11(0.01)	&-3.45(0.18)	&3.72(0.52)	&0.0279	&\\
71024.781	&H$_2^{13}$CO		&$1_{0,1}-0_{0,0}$		&0.24(0.01)	&-3.19(0.10)	&4.49(0.25)	&0.0499	&\\
71889.596	&HC$_5$N		&$27-26$			&0.31(0.01)	&-3.02(0.10)	&4.51(0.22)	&0.0640	&\\
\hline
\end{tabular}
\end{table}
\addtocounter{table}{-1}

\clearpage\newpage
\begin{table}
\caption{ (continued)
%\label{tab:colden}
}
\medskip
\footnotesize
\begin{tabular}{|l|l|l|llll|l|}
%\tablewidth{0pt}
\hline
Frequency & Molecule & Transition & $\int T_RdV$ & $V_{LSR}$ & $\Delta V$ & $T_R$ & Notes\\
 (MHz)  &          &        &(K km s$^{-1})$&(km s$^{-1}$)&(km s$^{-1}$ )& (K)   & \\
\hline
72039.331	&DCO$^+$		&$1-0$				&0.54(0.02)	&-2.99(0.05)	&3.56(0.13)	&0.1416	&\\
72107.721       &CCD                    &$1-0~J=5/2-3/2~F=7/2-5/2$      &0.26(0.02)     &-3.35(0.18)    &5.19(0.54)     &0.048  &\\
72187.708       &CCD                    &$1-0~J=3/2-3/2~F=5/2-5/2$      &0.08(0.01)     &-2.87(0.15)    &3.69(0.22)     &0.021  &new\\
72189.726       &CCD                    &$1-0~J=3/2-1/2~F=3/2-5/2$      &0.08(0.01)     &-2.87(0.15)    &3.69(0.22)     &0.020  &new\\
72323.789       &CCS                    &$J_N=6_5-5_4$                  &0.11(0.01)     &-3.25(0.14)    &3.41(0.18)     &0.031  &\\
72409.092	&H$_2$CO		&$5_{1,4}-5_{1,5}$		&0.28(0.01)	&-2.78(1.98)	&4.89(0.02)	&0.0544	&\\
72413.484	&DCN			&$1-0~F=1-1$			&0.75(0.01)	&-3.67(0.01)	&4.30(0.02)	&0.1752	&\\
72414.905	&DCN			&$1-0~F=2-1$			&1.26(0.01)	&-3.67(0.01)	&4.30(0.02)	&0.292	&\\
72417.030	&DCN			&$1-0~F=0-1$			&0.25(0.01)	&-3.67(0.01)	&4.30(0.02)	&0.0584	&\\
72475.074 	&HC$^{13}$CCN		&$8-7$				&0.18(0.01)	&-3.07(0.18)	&5.90(0.46)	&0.0293	&\\
72482.055 	&HCC$^{13}$CN		&$8-7$				&0.14(0.01)	&-3.02(0.19)	&4.63(0.38)	&0.0290	&\\
72758.235 	&SO$_2$			&$6_{0,6}-5_{1,5}$		&0.86(0.02)	&-3.31(0.04)	&4.77(0.11)	&0.1696	&rw\\ 
72783.818 	&HC$_3$N		&$8-7$				&6.52(0.02)	&-3.20(0.01)	&4.32(0.01)	&1.4191	&bw,rw\\
72835           &U                      &                               &0.11(0.01)     &-3.38(0.07)    &2.01(0.12)     &0.052  &new\\
72837.948 	&H$_2$CO		&$1_{0,1}-0_{0,0}$		&5.87(0.12)	&-4.19(0.03)	&2.97(0.04)	&1.86	&bw\\
                &                       &                               &3.43(0.12)     &-1.30(0.04)    &2.88(0.05)     &1.12   &\\
72976.779       &OCS			&$6-5$				&0.37(0.01)	&-3.05(0.07)	&4.10(0.15)	&0.0855	&\\
73577.454	&CH$_3$CN		&$4_3-3_3$			&0.11(0.01)	&-3.27(0.02)	&4.26(0.00)	&0.0244	&\\
73584.545	&CH$_3$CN		&$4_2-3_2$			&0.32(0.01)	&-3.27(0.02)	&4.26(0.00)	&0.0694	&\\
73588.801	&CH$_3$CN		&$4_1-3_1$			&0.81(0.01)	&-3.27(0.02)	&4.26(0.00)	&0.1787	&bl\\
73590.220	&CH$_3$CN		&$4_0-3_0$			&0.92(0.01)	&-3.27(0.02)	&4.26(0.03)	&0.2020	&bl\\
73722.376	&CH$_3$OCH$_3$		&$9_{2,7}-9_{1,8}EE$    	&0.14(0.02)	&-5.12(0.68)	&11.79(1.23)	&0.0109	&\\
+73720.490	&CH$_3$OCH$_3$		&$9_{2,7}-9_{1,8}AE+EA$	        &               &               &               &       &\\
+73724.261	&CH$_3$OCH$_3$		&$9_{2,7}-9_{1,8}AA$   	        &               &               &               &       &\\
%73839.235	&CH$_3$OH		&$9_{1,8}-10_{2,9}A^-~vt=1$	&0.07(0.01)	&-0.58(0.63)	&6.70(0.10)	&0.0096	&\\
74551.988	&HC$_5$N		&$28-27$			&0.28(0.01)	&-3.06(1.37)	&4.35(0.08)	&0.0594	&\\
74891.681	&CH$_3$CHO		&$4_{1,4}-3_{1,3}A^{++}$	&0.25(0.01)	&-3.27(1.87)	&4.35(0.18)	&0.0536	&\\
\hline
\end{tabular}
\end{table}

\addtocounter{table}{-1}
\clearpage\newpage
\begin{table}
\caption{ (continued)
%\label{tab:colden}
}
\medskip
\footnotesize
\begin{tabular}{|l|l|l|llll|l|}
%\tablewidth{0pt}
\hline
Frequency & Molecule & Transition & $\int T_RdV$ & $V_{LSR}$ & $\Delta V$ & $T_R$ & Notes\\
 (MHz)  &          &        &(K km s$^{-1})$&(km s$^{-1}$)&(km s$^{-1}$ )& (K)   & \\
\hline

74924.137	&CH$_3$CHO		&$4_{-1,4} - 3_{-1,3} E$	&0.22(0.01)	&-3.27(2.41)	&3.59(0.23)	&0.0571	&\\
75366           &U                      &                               &0.05(0.013)	&-3.27(0.23)	&2.66(0.58)	&0.018	&m\\
75862.92	&CH$_3$SH		&$3_0-2_0A+$			&0.16(0.02)	&-2.63(0.90)	&13.21(1.80)	&0.0115	&\\ +75864.43       &CH$_3$SH               &$3_0-2_0E$                     &               &               &               &       &\\
75921    	&U               	&        			&0.08(0.02)	&-3.57(0.93)	&8.05(1.67)	&0.001	&m\\
76117.43	&C$_4$H                 &17/2$-$15/2			&0.21(0.01)	&-3.02(0.08)	&3.10(0.19)	&0.0620	&\\
76156.02	&C$_4$H                 &15/2$-$13/2                    &0.20(0.01)	&-3.06(0.07)	&3.10(0.16)	&0.0613	&\\
76198.724	&l-C$_3$H		&$2P1/2 J=7/2-5/2,F=4-3f$	&0.089(0.009)	&-2.96(0.16)	&3.269(0.17)	&0.0257	&bl\\
76199.925	&l-C$_3$H		&$2P1/2 J=7/2-5/2,F=4-3f$	&0.072(0.008)	&-2.96(0.16)	&3.269(0.00)	&0.0206	&bl\\
76204.198	&l-C$_3$H		&$2P1/2 3-2 J=7/2-5/2~F=4-3$	&0.041(0.008)	&-2.96(0.16)	&3.269(0.00)	&0.0117	&\\
76205.108	&l-C$_3$H		&$2P1/2 3-2 J=7/2-5/2~F=3-2$	&0.046(0.008)	&-2.96(0.16)	&3.269(0.00)	&0.0132	&\\
76247.312	&CH$_3$OH		&$11_{1,10}-10_{2,9}A^-$	&0.12(0.01)	&-3.15(0.19)	&4.03(0.43)	&0.0282	&\\
76305.717	&DNC			&$1-0$				&0.91(0.01)	&-3.20(0.02)	&4.12(0.05)	&0.2066	&\\
76362.194	&CH$_3$OCH$_3$		&$7_{2,5}-7_{1,6} AE+EA$	&0.044(0.010)	&-3.43(0.35)	&4.56(0.00)	&0.0090	&m\\
76364.277	&CH$_3$OCH$_3$		&$7_{2,5}-7_{1,6} EE$		&0.067(0.010)	&-3.43(0.35)	&4.56(0.45)	&0.0137	&\\
%76366.367	&CH$_3$OCH$_3$		&$7_{2,5}-7_{1,6} AA$		&0.007(0.009)	&-3.43(0.35)	&4.56(0.00)	&0.0013	&\\
76412.158	&SO$_2$			&$10_{1,9}-9_{2,8}$		&0.27(0.01)	&-2.53(0.17)	&6.22(0.39)	&0.0402	&\\
76460           & U                     &                               &0.17(0.01)     &-5.12(0.04)    &1.32(0.12)     &0.120  &\\
                &                       &                               &0.20(0.02)     &-2.00(0.10)    &2.69(0.31)     &0.068  &\\
76509.628	&CH$_3$OH		&$5_{0,5}-4_{1,3} E$		&0.75(0.01)	&-2.99(0.04)	&5.40(0.09)	&0.1313	&\\
76866.437	&CH$_3$CHO		&$4_{0,4}-3_{0,3} E$	        &0.36(0.01)	&-3.21(0.06)	&4.29(0.14)	&0.0787	&\\
76878.958	&CH$_3$CHO		&$4_{0,4}-3_{0,3}A^{++}$	&0.36(0.01)	&-3.13(0.06)	&4.04(0.16)	&0.0832	&\\
77038.605	&CH$_3$CHO		&$4_{2,3}-3_{2,2}A^{--}$ 	&0.08(0.01)	&-3.62(0.20)	&3.50(0.46)	&0.0223	&\\
77107.86	&N$_2$D$^+$		&$1-0~F1=1-1$			&0.045(0.008)	&-3.16(0.16)	&3.20(0.00)	&0.0131	&\\
\hline
\end{tabular}
\end{table}

\addtocounter{table}{-1}
\clearpage\newpage
\begin{table}
\caption{ (continued)
%\label{tab:colden}
}
\medskip
\footnotesize
\begin{tabular}{|l|l|l|llll|l|}
%\tablewidth{0pt}
\hline
Frequency & Molecule & Transition & $\int T_RdV$ & $V_{LSR}$ & $\Delta V$ & $T_R$ & Notes\\
 (MHz)  &          &        &(K km s$^{-1})$&(km s$^{-1}$)&(km s$^{-1}$ )& (K)   & \\
\hline
77109.61	&N$_2$D$^+$		&$1-0~F1=2-1$			&0.074(0.008)	&-3.16(0.16)	&3.20(0.20)	&0.0215	&\\
77112.2		&N$_2$D$^+$     	&$1-0~F1=0-1$			&0.025(0.007)	&-3.16(0.16)	&3.20(0.00)	&0.0074	&hfs\\
77125.695	&CH$_3$CHO		&$4_{2,2}-3_{2,1}$		&0.26(0.04)	&-4.63(0.15)	&6.13(0.38)	&0.040	&\\
+77126.418	&CH$_3$CHO		&$4_{2,3}-3_{-2,2}$		&               &               &               &       &\\ 
77214.360	&HC$_5$N		&$29-28$			&0.28(0.01)	&-2.97(0.09)	&4.65(0.19)	&0.0556	&\\
77218.295	&CH$_3$CHO		&$4_{2,2}-3_{2,1}A^{++}$	&0.09(0.01)	&-3.05(0.32)	&5.35(0.78)	&0.0160	&\\
77731.725	&CCS			&$6,6-5,5$			&0.16(0.01)	&-2.95(0.13)	&4.27(0.31)	&0.0348	&\\
78603.670       &$^{30}$Si$^{18}$O      & $2-1$                         &0.047(0.01)    &-2.98(0.60)    &5.00(1.21)     &0.009 &m\\
79099.313	&CH$_3$CHO		&$4_{1,3}-3_{1,2} E$		&0.32(0.01)	&-3.34(0.05)	&4.28(0.12)	&0.0704	&\\
79150.172	&CH$_3$CHO		&$4_{1,3}-3_{1,2}$		&0.34(0.01)	&-3.37(0.07)	&4.86(0.28)	&0.0656	&\\
79350.476	&H$^{13}$CCCN		&$9-8$				&0.12(0.01)	&-2.98(0.17)	&4.51(0.33)	&0.0252	&\\
79876.711	&HC$_5$N		&$30-29$			&0.27(0.01)	&-2.70(0.18)	&5.12(0.13)	&0.0488	&\\
80076.644	&CH$_2$CO		&$4_{1,4}-3_{1,3}$		&0.25(0.01)	&-3.15(0.06)	&3.73(1.69)	&0.0637	&\\
80479.940	&CH$_2$CN		&$4-3~J=11/2-9/2	$	&0.04(0.006)	&-4.62(0.18)	&2.37(0.90)	&0.016	&m\\
80490           &U                      &                               &0.05(0.01)     &-4.89(0.23)    &2.97(0.48)     &0.015  &\\
80578.283	&HDO			&$1_{1,0}-1_{1,1}	$	&0.072(0.010)	&0.066(8.05)	&6.76(1.40)	&0.010	&di,bl?\\
80723.186	&c-C$_3$H$_2$		&$4_{2,2}-4_{1,3}	$	&0.046(0.009)	&-3.68(0.44)	&4.74(0.75)	&0.0091	&\\
80820.409	&CH$_2$CO		&$4_{2,3}-3_{2,2}	$	&0.030(0.007)	&-3.24(0.13)	&3.50(0.00)	&0.0082	&\\
80824.314	&CH$_2$CO		&$4_{2,2}-3_{2,1}	$	&0.024(0.007)	&-3.24(0.13)	&3.50(0.00)	&0.0063	&\\
80832.107	&CH$_2$CO		&$4_{0,4}-3_{0,3}	$	&0.11(0.01)	&-3.24(0.13)	&3.50(0.17)	&0.0283	&\\
80993.257	&CH$_3$OH		&$7_{2,6}-8_{1,7}A^-$	        &0.14(0.01)	&-2.62(2.04)	&6.62(3.10)	&0.0192	&\\
81477.49	&HNO			&$1_{0,1}-0_{0,0}	$	&0.12(0.01)	&-3.25(1.19)	&3.20(0.42)	&0.0344	&\\
81505.208	&CCS			&$7,6-6,5$			&0.31(0.01)	&-3.23(0.32)	&3.53(0.06)	&0.0834	&\\
81534.125	&HC$^{13}$CCN		&$9-8			$	&0.12(0.01)	&-2.68(1.60)	&4.74(0.19)	&0.0236	&\\
\hline
\end{tabular}
\end{table}

\addtocounter{table}{-1}
\clearpage\newpage
\begin{table}
\caption{ (continued)
%\label{tab:colden}
}
\medskip
\footnotesize
\begin{tabular}{|l|l|l|llll|l|}
%\tablewidth{0pt}
\hline
Frequency & Molecule & Transition & $\int T_RdV$ & $V_{LSR}$ & $\Delta V$ & $T_R$ & Notes\\
 (MHz)  &          &        &(K km s$^{-1})$&(km s$^{-1}$)&(km s$^{-1}$ )& (K)   & \\
\hline
81541.981	&HCC$^{13}$CN		&$9-8			$	&0.12(0.01)	&-2.66(1.73)	&3.77(1.13)	&0.0287	&\\
81586.229	&CH$_2$CO		&$4_{1,3}-3_{1,2}	$	&0.24(0.01)	&-3.00(0.42)	&3.67(3.11)	&0.0617	&\\
81881.462	&HC$_3$N		&$9-8			$	&6.22(0.01)	&-3.12(0.00)	&4.44(0.00)	&1.3159	&\\
82093.555	&c-C$_3$H$_2$		&$2_{0,2}-1_{1,1}	$	&0.47(0.01)	&-3.05(0.03)	&3.45(0.11)	&0.1266	&\\
82395.089       &l-C$_3$H$_2$           &$4_{1,4}-3_{1,3}$              &0.06(0.004)    &-2.59(0.07)    &2.20(0.76)     &0.0237 &\\
82539.375	&U  			&                               &0.15(0.01)	&-2.13(0.10)	&3.26(0.26)	&0.0440	&\\
         	&			&        			&0.049(0.008)	&0.79(0.10)	&1.71(0.22)	&0.0267 &\\
82966.201	&c-C$_3$H$_2$		&$3_{1,2}-3_{0,3}	$	&0.19(0.01)	&-2.89(0.08)	&3.84(0.17)	&0.0453	&\\   
83163		&U			&				&0.11(0.01)	&-3.78(0.14)	&3.99(0.31)	&0.0261	&\\
83207.51	&CH$_2$CHCN		&$9_{1,9}-8_{1,8}$		&0.021(0.004)	&-2.73(0.25)	&2.39(0.52)	&0.0084	&\\
83217           &U                      &                               &0.09(0.01)     &-4.03(0.13)    &3.64(0.31)     &0.024  &\\
83584.282	&CH$_3$CHO		&$2_{-1,2}-1_{0,1} E	$	&0.05(0.01)	&-2.88(0.23)	&3.15(0.48)	&0.0136	&\\
83688.086	&SO$_2$			&$8_{1,7}-8_{0,8}$		&0.80(0.01)	&-2.79(0.03)	&6.40(0.08)	&0.1179	&rw\\
83900.570       &HOCN                   &$4_{0,4}-3_{0,3}$		&0.04(0.01)	&-3.74(0.22)	&3.02(0.62)	&0.0139	&di\\
83933.681	&l-C$_3$H$_2$           &$4_{1,3}-3_{1,2}$              &0.06(0.01)	&-2.71(0.16)	&2.47(0.59)	&0.0211	&\\
84119.329       &$^{13}$CCH             &$1-0~3/2-1/2~F=2,2.5-1,1.5$    &0.07(0.01)     &-3.02(0.17)    &3.66(0.33)     &0.017  &\\
84410.693	&$^{34}$SO		&$N,J=2,2-1,1		$	&0.13(0.01)	&-0.92(0.37)	&9.28(0.94)	&0.0128	&bl\\ 
84423.706	&CH$_3$OH		&$13_{-3,11}-14_{-2,13}E$	&0.05(0.01)	&-3.76(0.15)	&2.69(0.35)	&0.0165	&\\ 
84429.815       &DC$_3$N                & $10-9$                        &0.08(0.01)     &-2.67(0.28)    &4.49(0.73)     &0.016  &\\
84449.102	&CH$_3$OCHO		&$7_{2,6}-6_{2,5} E	$	&0.03(0.01)	&-2.49(0.00)	&5.82(0.00)	&0.0054	&m\\ 
84454.787	&CH$_3$OCHO		&$7_{2,6}-6_{2,5} A	$	&0.04(0.01)	&-2.49(0.43)	&5.82(0.51)	&0.0065	&m\\ 
84521.206	&CH$_3$OH		&$5_{-1,5}-4_{0,4} E	$	&0.85(0.04)	&0.43(0.00)	&0.89(0.00)	&0.9033	&mas\\ 
         	&        		&                     		&0.005(0.005)	&0.16(0.01)	&0.76(0.01)	&0.0066	&mas\\ 
         	&        		&                     		&2.27(0.11)	&-0.70(0.05)	&2.62(0.16)	&0.8144	&mas\\ 
         	&        		&                    		&3.65(0.12)	&-3.84(0.05)	&3.77(0.21)	&0.9108	&mas\\ 
\hline
\end{tabular}
\end{table}

\addtocounter{table}{-1}
\clearpage\newpage
\begin{table}
\caption{ (continued)
%\label{tab:colden}
}
\medskip
\footnotesize
\begin{tabular}{|l|l|l|llll|l|}
%\tablewidth{0pt}
\hline
Frequency & Molecule & Transition & $\int T_RdV$ & $V_{LSR}$ & $\Delta V$ & $T_R$ & Notes\\
 (MHz)  &          &        &(K km s$^{-1})$&(km s$^{-1}$)&(km s$^{-1}$ )& (K)   & \\
\hline
84542.331	&NH$_2$CHO		&$4_{0,4}-3_{0,3}$		&0.031(0.006)	&-3.55(0.27)	&3.21(0.75)	&0.0091	&\\
84727.691	&c-C$_3$H$_2$		&$3_{2,2}-3_{1,3}$		&0.07(0.01)	&-3.39(0.23)	&4.14(0.98)	&0.0161	&\\
84745.998	&$^{30}$SiO		&$2-1~v=0	$		&0.15(0.01)	&-3.97(0.21)	&7.21(0.79)	&0.0198	&\\
84865.166       & O$^{13}$CS            &$7-6$                          &0.03(0.008)    &-4.73(0.78)    & 5.68(1.40)    & 0.006&m\\
85139.104	&OCS			&$7-6		$		&0.55(0.01)	&-2.86(0.03)	&4.63(0.07)	&0.1108	&\\
85153.931	&D$_2$CS		&$3_{0,3}-2_{0,2}$		&0.023(0.007)	&-3.46(0.40)	&3.00(1.14)	&0.0073	&di\\
85162.223	&HC$^{18}$O$^+$		&$1-0$				&0.43(0.01)	&-3.35(0.03)	&3.93(0.07)	&0.1016	&\\
85201.346	&HC$_5$N		&$32-31$        		&0.25(0.02)	&-3.89(0.13)	&2.72(0.26)	&0.089	&\\ 
         	&       		&                		&0.14(0.02)	&-0.93(0.22)	&2.55(0.38)	&0.053	&\\ 
85229.27	&C$^{13}$CH		&$1-0~3/2-1/2~F=2,2.5-1,1.5$	&0.09(0.01)	&-2.93(0.11)	&4.26(0.16)	&0.0205	&\\ 
85232.76	&C$^{13}$CH		&$1-0~3/2-1/2~F=2,1.5-1,0.5$	&0.07(0.01)	&-2.93(0.11)	&4.26(0.00)	&0.0149	&\\ 
85247.65	&C$^{13}$CH		&$1-0~3/2-1/2~F=1,0.5-0,0.5$	&0.035(0.006)	&-2.93(0.11)	&4.26(0.00)	&0.0078	&\\ 
85256.96	&C$^{13}$CH		&$1-0~3/2-1/2~F=1,1.5-0,0.5$	&0.06(0.01)	&-2.93(0.11)	&4.26(0.00)	&0.0133	&\\ 
85265.507	&t-CH$_3$CH$_2$OH	&$6_{0,6}-5_{1,5}	$	&0.06(0.01)	&-2.12(0.41)	&5.89(0.53)	&0.0094	&\\ 
85307.459       &C$^{13}$CH             &$1-0~1/2-1/2~F=1,1.5-1,1.5$    &0.06(0.006)    &-2.62(0.23)    &4.20(0.44)    &0.014  &\\
85338.906	&c-C$_3$H$_2$		&$2_{1,2}-1_{0,1}	$	&1.51(0.01)	&-3.08(0.01)	&3.72(0.01)	&0.3807	&\\ 
85347.869	&HCS$^+$		&$2-1			$	&0.71(0.01)	&-3.01(0.09)	&3.93(0.01)	&0.1694	&\\ 
85442.6		&CH$_3$CCH		&$5_{3}-4_{3}		$	&0.26(0.00)	&-3.24(0.01)	&4.09(0.00)	&0.0587	&\\ 
85450.765	&CH$_3$CCH		&$5_{2}-4_{2}		$	&0.43(0.00)	&-3.24(0.01)	&4.09(0.00)	&0.0978	&\\ 
85455.665	&CH$_3$CCH		&$5_{1}-4_{1}		$	&1.02(0.00)	&-3.24(0.01)	&4.09(0.00)	&0.2348	&\\ 
85457.299	&CH$_3$CCH		&$5_{0}-4_{0}		$	&1.27(0.01)	&-3.24(0.01)	&4.09(0.01)	&0.2910	&\\ 
85531.512       &HOCO$^+$               &$4_{0,4}-4_{1,3}$              &0.03(0.006)    &-3.41(0.28)    &2.78(0.63)     &0.011  &m\\ 
85568.074	&CH$_3$OH		&$6_{-2,5}-7_{-1,7} E	$	&0.10(0.01)	&-3.41(0.14)	&4.08(0.37)	&0.0238	&\\ 
\hline
\end{tabular}
\end{table}

\addtocounter{table}{-1}
\clearpage\newpage
\begin{table}
\caption{ (continued)
%\label{tab:colden}
}
\medskip
\footnotesize
\begin{tabular}{|l|l|l|llll|l|}
%\tablewidth{0pt}
\hline
Frequency & Molecule & Transition & $\int T_RdV$ & $V_{LSR}$ & $\Delta V$ & $T_R$ & Notes\\
 (MHz)  &          &        &(K km s$^{-1})$&(km s$^{-1}$)&(km s$^{-1}$ )& (K)   & \\
\hline
85634.00        &C$_4$H			&$9-8~J=19/2-17/2	$	&0.17(0.01)	&-2.85(0.03)	&2.60(0.04)	&0.0631	&\\ 
85672.57	&C$_4$H			&$9-8~J=17/2-15/2	$	&0.16(0.01)	&-2.85(0.03)	&2.60(0.00)	&0.0561	&\\ 
85656.418	&c-C$_3$H$_2$		&$4_{3,2}-4_{2,3}	$	&0.039(0.005)	&-3.67(0.18)	&3.20(0.75)	&0.0113	&\\ 
85715.424	&CH$_2$CHCN		&$9_{2,7}-8_{2,6}	$	&0.03(0.006)	&-3.42(0.28)	&2.71(0.50)	&0.010	&\\ 
85759.188	&$^{29}$SiO		&$2-1 v=0		$	&0.20(0.01)	&-3.29(0.11)	&6.44(0.01)	&0.0292	&\\
85926.270       &NH$_2$D                &$1_{1,1}-1_{0,1}$              &0.99(0.02)     &-2.94(0.06)    &7.09(0.15)     &0.131  &hfs\\
86054.967	&HC$^{15}$N		&$1-0			$	&1.23(0.01)	&-3.02(0.02)	&4.60(0.01)	&0.2512	&\\ 
86093.983	&SO			&$N,J=2,2-1,1$			&2.27(0.01)	&-2.59(0.01)	&5.33(0.03)	&0.3997	&\\
86181.413	&CCS			&$N,J=7,6-6,5$			&0.13(0.01)	&-2.98(0.13)	&3.72(0.33)	&0.0337	&\\
86338.736	&H$^{13}$CN		&$1-0~F=1-1$			&1.95(0.01)	&-3.38(0.00)	&4.56(0.00)	&0.4028	&\\
86340.176	&H$^{13}$CN		&$1-0~F=2-1$			&3.06(0.01)	&-3.38(0.00)	&4.56(0.01)	&0.6308	&\\
86342.255	&H$^{13}$CN		&$1-0~F=0-1$			&0.71(0.01)	&-3.38(0.00)	&4.56(0.00)	&0.1474	&\\
86546.18	&t-HCOOH		&$4_{1,4}-3_{1,3}$		&0.10(0.01)	&-3.11(0.22)	&4.56(0.74)	&0.0214	&\\
86615.602	&CH$_3$OH		&$7_{2,6}-6_{3,3}A^{--}$	&0.11(0.01)	&-3.17(0.23)	&3.67(0.57)	&0.0283	&\\
86670.82	&HCO			&$1_{0,1}-0_{0,0}~3/2-1/2~F=2-1$&0.51(0.00)	&-2.53(0.02)	&2.87(0.03)	&0.1667	&\\
86708.35	&HCO			&$1_{0,1}-0_{0,0}~3/2-1/2~F=1-0$&0.36(0.01)	&-2.53(0.02)	&2.87(0.00)	&0.1172	&\\
86754.288	&H$^{13}$CO$^+$		&$1-0$				&4.10(0.01)	&-3.35(0.00)	&4.09(0.00)	&0.9428	&\\
86777.43	&HCO			&$1_{0,1}-0_{0,0}~1/2-1/2~F=1-1$&0.31(0.01)	&-2.85(0.04)	&2.92(0.07)	&0.1009	&\\
86805.75	&HCO			&$1_{0,1}-0_{0,0}~1/2-1/2~F=0-1$&0.13(0.01)	&-2.85(0.04)	&2.92(0.00)	&0.0409	&\\
86846.995	&SiO			&$2-1~v=0$			&1.70(0.02)	&-3.27(0.26)	&4.59(0.26)	&0.3483	&\\
         	&                       &				&1.52(0.02)	&-3.40(0.26)	&16.60(0.26)	&0.0859	&\\
86902.947	&CH$_3$OH		&$7_{2,5}-6_{3,4}A^{++}$	&0.09(0.01)	&-3.34(0.15)	&3.48(0.67)	&0.0236	&\\
87057.258	&HC$^{17}$O$^+$		&$1-0~F=7/2-5/2$		&0.14(0.01)	&-3.21(0.30)	&6.47(0.80)	&0.0200	&\\
+87056.966      &HC$^{17}$O$^+$		&$1-0~F=3/2-5/2$		&               &               &               &       &\\
+87058.294      &HC$^{17}$O$^+$		&$1-0~F=5/2-5/2$		&               &               &               &       &\\
87090.859	&HN$^{13}$C		&$1-0~F=2-1$			&1.73(0.01)	&-2.97(0.01)	&4.32(0.03)	&0.3771	&\\
87143.198	&CH$_3$OCHO		&$7_{3,4}-6_{3,3}E$		&0.09(0.01)	&-3.74(0.37)	&5.66(0.76)	&0.0141	&\\%bl. C4H
87163   	& U                     &                        	&0.05(0.007)	&-3.42(0.14)	&1.95(0.38)	&0.024	&\\
\hline
\end{tabular}
\end{table}

\addtocounter{table}{-1}
\clearpage\newpage
\begin{table}
\caption{ (continued)
%\label{tab:colden}
}
\medskip
\footnotesize
\begin{tabular}{|l|l|l|llll|l|}
%\tablewidth{0pt}
\hline
Frequency & Molecule & Transition & $\int T_RdV$ & $V_{LSR}$ & $\Delta V$ & $T_R$ & Notes\\
 (MHz)  &          &        &(K km s$^{-1})$&(km s$^{-1}$)&(km s$^{-1}$ )& (K)   & \\
\hline
87284.156	&C$_2$H			&$1-0~3/2-1/2~F=1-1$		&1.03(0.00)	&-2.78(0.02)	&3.70(0.01)	&0.2624	&\\
87313   	& U      		&                		&0.17(0.01)	&-5.11(0.09)	&2.97(0.12)	&0.0539	&\\
87316.925	&C$_2$H			&$1-0~3/2-1/2~F=2-1$		&7.10(0.01)	&-3.01(0.00)	&4.23(0.00)	&1.5753	&\\
87328.624	&C$_2$H			&$1-0~3/2-1/2~F=1-0$		&4.19(0.01)	&-3.01(0.00)	&4.23(0.00)	&0.9312	&\\
87398           &U                      &                               &0.04(0.01)     &-4.73(0.21)    &1.70(0.43)     &0.020  &\\
87402.004	&C$_2$H			&$1-0~1/2-1/2~F=1-1$		&4.32(0.01)	&-3.01(0.00)	&4.23(0.00)	&0.9580	&\\
87407.165	&C$_2$H			&$1-0~1/2-1/2~F=0-1$		&2.14(0.01)	&-3.01(0.00)	&4.23(0.00)	&0.4748	&\\
87446.512	&C$_2$H			&$1-0~1/2-1/2~F=1-0$		&1.18(0.01)	&-3.01(0.00)	&4.23(0.00)	&0.2615	&\\
87863.63	&HC$_5$N		&$33-32$			&0.20(0.01)	&-2.83(0.12)	&5.01(1.11)	&0.0381	&\\
\hline
\end{tabular}

\medskip\flushleft
Notes: m -- detection at the sensitivity limit; bl -- line blends with its neighbours; bw -- line has a blue wing; rw -- line has a red wing; di -- doubtful identification; hfs -- line has an hyperfine structure; new -- line does not exist in the Lovas database~\cite{lovas09}; mas -- maser line;
\end{table}

\end{document}